\documentclass[conference]{IEEEtran}
\IEEEoverridecommandlockouts

\newif\ifdraft
\drafttrue 

\usepackage{amsmath, amssymb}   
\usepackage{algorithm}          
\usepackage{algorithmic}        
\usepackage{booktabs}
\usepackage{graphicx}
\usepackage{subcaption}

\ifdraft
\newcommand{\kavindu}[1]{\todo[inline,color=brown!40]{kavindu: #1}}
\newcommand{\josh}[1]{\todo[inline,color=blue!40]{josh: #1}}
\newcommand{\saurabh}[1]{\todo[inline,color=red!40]{saurabh: #1}}
\newcommand{\rev}[1]{\textcolor{black}{#1}}
\else
\newcommand{\kavindu}[1]{}
\newcommand{\josh}[1]{}
\newcommand{\saurabh}[1]{}
\newcommand{\rev}[1]{#1}
\fi

\usepackage[letterpaper, margin=1in]{geometry}
\usepackage{cite}
\usepackage{amsmath,amssymb,amsfonts}
\usepackage{algorithmic}
\usepackage{graphicx}
\usepackage{textcomp}
\usepackage{xcolor}

\def\BibTeX{{\rm B\kern-.05em{\sc i\kern-.025em b}\kern-.08em
    T\kern-.1667em\lower.7ex\hbox{E}\kern-.125emX}}

\usepackage{todonotes}

\usepackage{fullpage}
\usepackage{url}
\usepackage{wrapfig}
\usepackage[font={bf,it,small},labelsep=colon,justification=centering]{caption}
\usepackage{tikz}
\usetikzlibrary{patterns}
\usepackage{pgfplots}
\usepackage{graphicx}
\usepackage{amsmath}
\usepackage{setspace}
\usepackage{multirow}
\usepackage{color}
\usepackage{subcaption}
\usepackage{xparse}
\usepackage{expl3}

\usepackage{algorithm} 
\usepackage{algorithmic} 
\usepackage{mathtools} 
\usepackage{amsthm, amssymb, amscd, amsfonts} 
\usepackage{bm} 
\usepackage{paralist}

\renewcommand{\thepage}{\arabic{page}}

\usepackage{enumitem}
\usepackage{xcolor,xspace}

\usepackage{hyperref}
\hypersetup{
  colorlinks=true, 
  linktoc=all,     
  linkcolor=blue,  
  citecolor=blue,
  urlcolor = black,
}

\usepackage{pdfpages}

\usepackage{csquotes}
\usepackage{comment}
\usepackage[hang,flushmargin,perpage]{footmisc}

\usepackage[most]{tcolorbox}
\newtcolorbox[%
auto counter]{mybox}[2][]{%
	enhanced jigsaw,
	breakable,
	opacityback=0, 
	#1}

\newcommand{\name}{{\sc SABLE}\xspace}

\begin{document}

\title{Beyond Corner Patches: Semantics-Aware Backdoor Attack in Federated Learning \\
{\footnotesize \textsuperscript{*}Accepted as a regular paper at IEEE/IFIP International Conference on Dependable Systems and Networks (DSN), 2026}
}

\author{
    \IEEEauthorblockN{
        Kavindu Herath,
        Joshua Zhao,
        Saurabh Bagchi
    }
    \IEEEauthorblockA{
        \{kherathm, zhao1207, sbagchi\}@purdue.edu
    }
    \IEEEauthorblockA{
        Department of Electrical and Computer Engineering, Purdue University, \\ West Lafayette, USA
    }
    
}

\maketitle

\begin{abstract}

Backdoor attacks on federated learning (FL) are most often evaluated with synthetic corner patches or out-of-distribution (OOD) patterns that are unlikely to arise in practice. In this paper, we revisit the backdoor threat to standard FL (a single global model) under a more realistic setting where triggers must be semantically meaningful, in-distribution, and visually plausible. We propose \emph{\name}, a \textbf{S}emantics-\textbf{A}ware \textbf{B}ackdoor for \textbf{LE}arning in federated settings, which constructs natural, content-consistent triggers (e.g., semantic attribute changes such as sunglasses) and optimizes an aggregation-aware malicious objective with feature separation and parameter regularization to keep attacker updates close to benign ones. We instantiate \emph{\name} on CelebA hair-color classification and the German Traffic Sign Recognition Benchmark (GTSRB), poisoning only a small, interpretable subset of each malicious client's local data while otherwise following the standard FL protocol. Across heterogeneous client partitions and multiple aggregation rules (FedAvg, Trimmed Mean, MultiKrum, FLAME and FilterFL), our semantics-driven triggers achieve high targeted attack success rates while preserving benign test accuracy. These results show that semantics-aligned backdoors remain a potent and practical threat in federated learning, and that robustness claims based solely on synthetic patch triggers can be overly optimistic.

\end{abstract}


\begin{IEEEkeywords}
Federated learning, Backdoor attacks, Semantic triggers, Model poisoning, Robust aggregation
\end{IEEEkeywords}

\section{Introduction}

 
Federated learning (FL) has emerged as a popular paradigm for training deep models across decentralized datasets without aggregating raw data at a central server~\cite{mcmahan2017communication,kairouz2021advances,rahman2021survey,zhao2025federation,zhao2024leak}.
It underpins applications such as mobile keyboard prediction, personalized vision models, and privacy-aware analytics, where data naturally resides on user devices or edge nodes~\cite{konecny2016federated,hard2019federated,zhao2023resource,zhao2024loki}.
At the same time, FL deployments must contend with heterogeneous, non-IID client data and intermittent participation~\cite{li2020federated}, while a small number of malicious clients can tamper with their local training data or updates and indirectly manipulate the global model~\cite{bagdasaryan2018backdoor,bhagoji2019model,wang2020attack}.
Among such threats, \emph{backdoor attack} is particularly insidious because it preserves aggregate accuracy while introducing targeted, hard-to-detect failures~\cite{gu2017badnets,chen2017targeted,liu2018trojaning}. In it, the malicious nodes pollute the global model so that it behaves normally on clean inputs but misclassifies any input containing a secret trigger to an attacker-chosen target class. \rev{Most commonly, this is done by training with a small attacker-chosen trigger appended to a natural image (often a small patch or shape added to the corner).}


\rev{Although backdoor attacks were originally introduced in traditional centralized learning~\cite{gu2017badnets, chen2017targeted}, recent works have also investigated the effectiveness of backdoors and model-poisoning attacks in FL. These methods explore increasing effectiveness through aggressive scaling of local model, backdooring under FL client sampling and non-IID data, and colluding clients~\cite{bagdasaryan2018backdoor, bhagoji2019model, xie2020dba}. As a result, many robust aggregation and Byzantine-resilient distributed learning approaches have also been designed to counteract the effect of the malicious client updates using filtering through distance-based or coordinate-wise checks~\cite{blanchard2017krum, yin2018byzantine}. Intuitively, one might hope that applying such robust aggregation against simple patch triggers would suffice: 
if an attacker pushes too hard on the backdoor, their update would be filtered; if they push gently, the backdoor would not be effective. 
}

\begin{figure*}[t]
    \centering
    \includegraphics[width=1\linewidth]{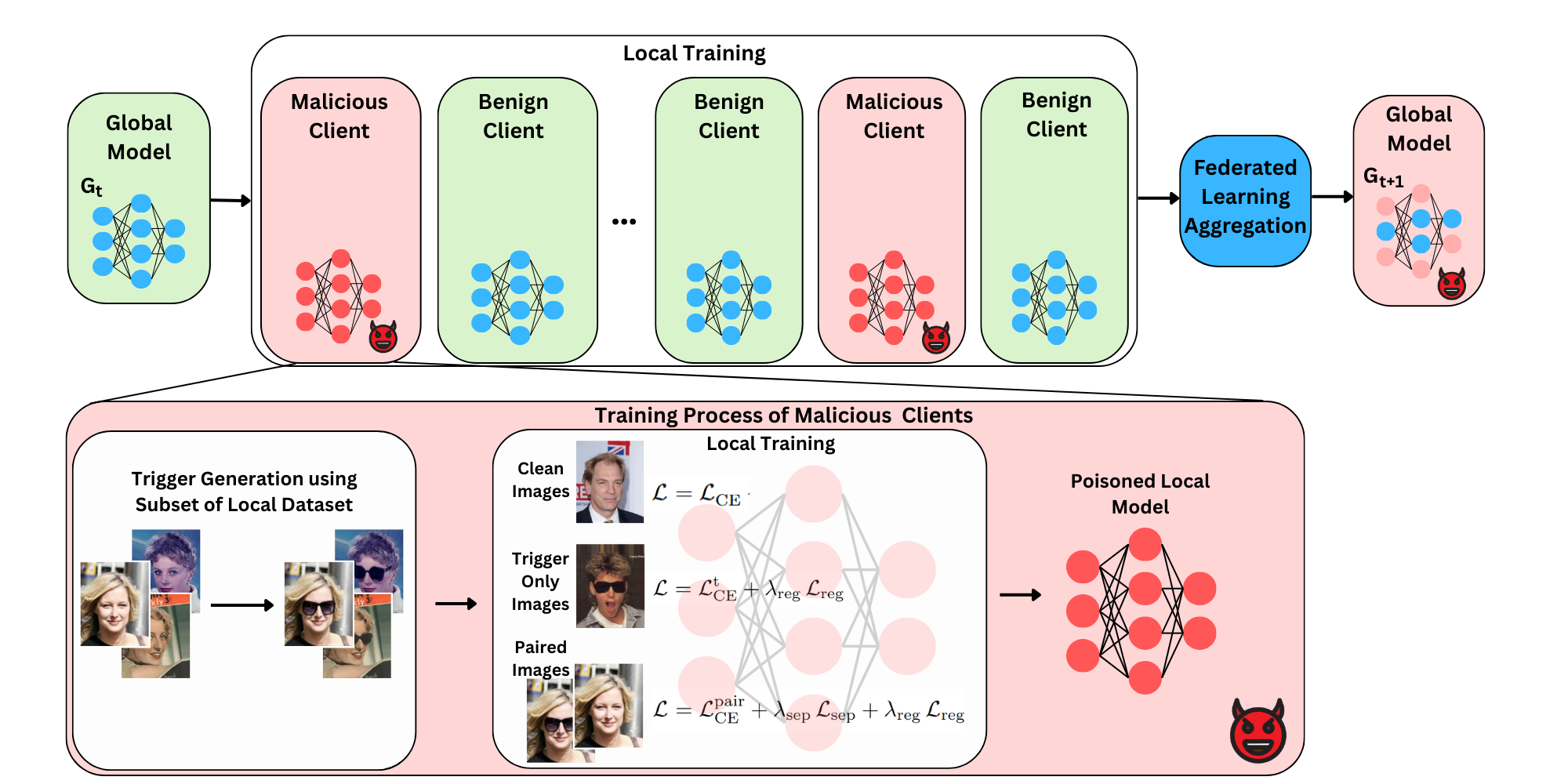}
    \caption{Overview of \name, our semantics-aware backdoor attack in federated learning. 
\textbf{Top:} Standard FL round with a mixture of benign and malicious clients. 
Benign clients train on clean local data, while malicious clients return poisoned updates that are aggregated by the server into the next global model. 
\textbf{Bottom:} Training pipeline on a malicious client. A subset of local images is used to generate semantic triggers, producing paired clean/triggered samples. 
The malicious client optimizes a joint objective that combines clean cross-entropy, triggered cross-entropy toward the target label, feature-separation loss on paired samples, and a regularization term that keeps its update close to the global model, yielding a poisoned local model that still appears benign under aggregation.}

    \label{fig:my_plot}
\end{figure*}

However, existing FL backdoor attacks and defenses share two key limitations.
First, most FL evaluations focus on \emph{synthetic}, patch-style triggers that are clearly out-of-distribution relative to the underlying content~\cite{gu2017badnets,bagdasaryan2018backdoor,xie2020dba,wang2020attack}.
Such triggers are easier to isolate via clustering or distance-based aggregation and do not capture realistic, semantics-preserving perturbations (for example, adding sunglasses that genuinely correspond to an object in the scene, or placing a small object near a traffic sign).
Only a few works systematically study semantic or physical backdoors, typically in centralized settings~\cite{saha2020hidden,wenger2021backdoor}, leaving open how such triggers behave under federated aggregation and non-IID client data.
Second, most FL backdoor attacks are not explicitly \emph{aggregation-aware}: they do not shape their local objectives to balance backdoor strength, benign accuracy, and similarity to benign updates under robust aggregation rules such as MultiKrum or Trimmed Mean~\cite{blanchard2017krum,yin2018byzantine}.
As a result, under strong defenses they often either lose the backdoor or incur a large benign accuracy drop by pushing easily detectable deviations from the global model~\cite{bagdasaryan2018backdoor,bhagoji2019model,wang2020attack}.

This paper asks: \emph{Can we construct a semantics-aware, physically realizable backdoor in FL that remains effective even under robust aggregation, while preserving benign accuracy?}
To this end, we introduce \emph{\name}, a \textbf{S}emantics-\textbf{A}ware \textbf{B}ackdoor for \textbf{LE}arning in Federated Learning settings.
\name couples \emph{natural, semantics-aligned triggers} with an \emph{aggregation-aware malicious objective} that explicitly separates clean and triggered features while regularizing the attacker update to stay close to the global model.
Instead of overlaying artificial patches, we generate triggers by semantically modifying base images so that the trigger aligns with the underlying object or region and remains visually plausible.
On the optimization side, \name exploits access to paired clean/triggered samples on malicious clients and uses a loss that preserves clean performance via standard cross-entropy, enforces a targeted backdoor on triggered samples, separates penultimate-layer representations of clean and triggered versions of the same input, and constrains the distance from the received global parameters so that the attacker’s update remains aggregation-friendly.
We further integrate a Neurotoxin-style gradient masking step that estimates parameter importance on clean data and selectively attenuates updates on the most critical parameters, encouraging the backdoor to reside in a low-importance subspace.

We instantiate \name for image classification FL with a small number of colluding malicious clients \rev{(e.g., 10--20\%)}.
Each malicious client maintains three disjoint subsets of local data (clean-only, trigger-only, and paired clean/triggered samples) and optimizes the joint malicious objective over these subsets, while the server runs standard FedAvg or robust aggregation, such as MultiKrum and Trimmed Mean, without modification.
We evaluate on CelebA and GTSRB, fixing a single target class and measuring global clean accuracy and attack success rate (ASR) under different aggregation rules.
We compare \name against a Bagdasaryan-style 
backdoor attack as baseline~\cite{bagdasaryan2018backdoor} that performs local backdoor training on mixed clean/triggered data and participates in FedAvg, Trimmed Mean, or MultiKrum without any feature separation, parameter regularization, or gradient masking.
Across all settings, \name consistently achieves high ASR while keeping clean accuracy close to benign training, and remains substantially more effective than the na\"ive baseline under all experimented aggregation methods. 

In summary, our contributions are as follows:
\begin{itemize}[leftmargin=*]
    \item We formulate and study \emph{\name}, a semantics-aware backdoor attack for federated learning in which triggers are visually plausible, physically realizable modifications rather than artificial pixel patches.
    \item We propose an aggregation-aware malicious objective for \name that combines clean and triggered cross-entropy, feature-space separation between paired clean/triggered samples, and parameter regularization to keep attacker updates close to the global model, further enhanced by Neurotoxin-style gradient masking on high-importance parameters.
    \item We instantiate \name on CelebA and GTSRB under standard FedAvg and robust aggregation (MultiKrum and Trimmed Mean), and empirically show that it substantially outperforms a Bagdasaryan-style naive backdoor baseline using local backdoor training without any regularization, even under strong defenses.
\end{itemize}

\section{Background and Related Work}
\label{sec:background}

\subsection{Federated Learning}

Federated learning (FL) considers a central server coordinating training across a population of clients, each holding a private local dataset~\cite{mcmahan2017communication,kairouz2021advances}.
Let $\mathcal{K} = \{1,\dots,K\}$ denote the set of clients, with local data distribution $\mathcal{D}_k$ and empirical loss
\begin{equation}
    F_k(w)
    \;=\;
    \mathbb{E}_{(x,y)\sim \mathcal{D}_k}\big[\ell(f_w(x),y)\big],
\end{equation}
where $f_w$ is the model with parameters $w$ and $\ell(\cdot,\cdot)$ is a standard supervised loss (e.g., cross-entropy).

The global FL objective is the weighted sum of local objectives,
\begin{equation}
    F(w)
    \;=\;
    \sum_{k\in\mathcal{K}} p_k F_k(w),
    \qquad
    p_k \;\propto\; |\mathcal{D}_k|,
    \label{eq:fl-objective}
\end{equation}
which is minimized by iterating through multiple rounds of server–client communication.
In the classical FedAvg algorithm~\cite{mcmahan2017communication}, round $t$ proceeds as follows.
The server broadcasts the current global model $w_t$ to a subset $\mathcal{S}_t \subseteq \mathcal{K}$ of clients.
Each participating client $k \in \mathcal{S}_t$ initializes its local model at $w_t$ and performs several steps of stochastic gradient descent on its local objective $F_k$, yielding an updated model $w_t^{(k)}$.
The server then aggregates these local models by a weighted average,
\begin{equation}
    w_{t+1}
    \;=\;
    \sum_{k \in \mathcal{S}_t} p_k \, w_t^{(k)},
    \label{eq:fedavg-update}
\end{equation}
which is a stochastic approximation to gradient descent on the global objective~\eqref{eq:fl-objective}.

\subsection{Backdoor Attacks}

A \emph{backdoor} (or trojan) attack implants a hidden behavior into a model: performance on clean inputs remains high, but the model is forced to output an erroneous, attacker-chosen target label $y^\star$ whenever a secret trigger pattern $\tau$ is present in the input~\cite{gu2017badnets,chen2017targeted,liu2018trojaning}.
Effectiveness is typically measured by (i) \emph{clean accuracy} on benign test data and (ii) \emph{attack success rate} (ASR), the fraction of triggered test inputs misclassified as the target label $y^\star$~\cite{chou2023backdoor, schwarzschild2021just, wu2022backdoorbench}.
This behavior can be induced by poisoning a small fraction of the training data with $(x\oplus\tau, y^\star)$ while keeping the rest clean~\cite{gu2017badnets,turner2019label,saha2020hidden,wenger2021backdoor}.

\rev{Backdoor attacks have been extensively studied in centralized training~\cite{li2022backdoor, liu2020reflection, doan2021lira}.
Early work such as BadNets showed that poisoning a small fraction of training data with images stamped by a fixed patch can induce high attack success rates (ASR) with negligible impact on clean accuracy~\cite{gu2017badnets}.
Subsequent work explored data-poisoning based backdoors~\cite{chen2017targeted}, model-supply-chain trojans~\cite{liu2018trojaning}, and label-consistent and hidden triggers~\cite{turner2019label,saha2020hidden,li2021invisible}.
Recent centralized works also demonstrate that \emph{semantic} or \emph{physical} triggers, such as eyeglasses or scarves worn by a person, can induce realistic backdoors that survive viewpoint changes and imaging noise~\cite{saha2020hidden,wenger2021backdoor}.
However, the behavior of these semantic and physical attacks have largely been unexplored 
in federated learning settings with robust aggregation.}

\subsection{Backdoor Attacks in FL}

\rev{In federated learning, backdoor attacks are realized as \emph{model poisoning}:
a malicious client modifies its local training objective and/or data so that its update both preserves benign accuracy and enforces the backdoor~\cite{bagdasaryan2018backdoor,bhagoji2019model,wang2020attack}. Compared to centralized settings, executing a backdoor attack in federated learning is \textit{more challenging} due to the uncertainty surrounding the server's aggregation protocol and the stochastic nature of client selection.}

\rev{Properties of backdoor and model-poisoning attacks unique to federated learning have also been investigated in several recent works. For example, a single malicious client can perform model replacement by aggressively training on backdoored data and scaling its local model to overwrite the global parameters under FedAvg~\cite{bagdasaryan2018backdoor}. Other examples include explicitly trading off benign accuracy and targeted misclassification within model-poisoning attacks~\cite{bhagoji2019model} or revisiting the feasibility of backdooring federated learning under client sampling and non-IID and heterogeneous data~\cite{sun2019can,wang2020attack}. In another line of work, distributed backdoor attacks (DBA) were proposed, where multiple colluding clients each embed a piece of a composite trigger, so that the global model learns a joint backdoor without any single client ever using the full pattern~\cite{xie2020dba}. Surveys such as~\cite{kairouz2021advances, zhang2022backdoorsurvey} synthesize these developments and highlight open robustness challenges. \textit{Despite this, most existing attacks still use synthetic patch-style triggers that are clearly out-of-distribution (e.g., corner squares, stickers, or high-contrast patterns)~\cite{gu2017badnets, bagdasaryan2018backdoor, xie2020dba}.}}

\subsection{Robust Aggregation and Backdoor Defense in Federated Learning}

Parallel to work federated learning backdoors and model-poisoning is robust aggregation and Byzantine-resilient distributed learning. Because vanilla FedAvg is highly sensitive to even a small number of corrupted updates (as it employs a magnitude-agnostic weighted average and doesn't enforce any outlier detection), a large body of work studies \textit{Byzantine-robust} aggregation techniques. These methods aim to dampen malicious updates using distance-based filters, coordinate-wise statistics, or consistency checks.
\rev{Early defenses such as Krum and MultiKrum~\cite{blanchard2017krum} select one or several client updates whose parameter vectors are closest (in squared Euclidean distance) to the rest, discarding those that appear as outliers. Coordinate-wise Median and Trimmed Mean~\cite{yin2018byzantine} operate per parameter: for each dimension, they either take the median or discard extreme values before averaging, thereby bounding the influence of arbitrary outliers.}


\rev{As this line of work progressed, vulnerability analyses such as El Mhamdi \textit{et al.}~\cite{mhamdi2018hidden} further highlighted the need for stronger defenses by showing that naive averaging can be arbitrarily misled even when most workers are honest. More recent defenses include FLAME~\cite{nguyen2022flame}, which combines robust aggregation with clustering-based filtering by first denoising client updates and then adaptively reweighting or discarding suspicious ones to mitigate backdoor attacks; FLAIR~\cite{sharma2023flair}, which defends against directed-deviation attacks through flip-score-based filtering; and FLTrust~\cite{cao2020fltrust}, which uses a small server-held dataset for trust bootstrapping. These and related methods~\cite{mhamdi2018hidden,zhao2022fedinv,zhao2021sear,fang2024byzantine} provide robustness guarantees when the number of Byzantine clients is bounded and have become standard baselines for secure federated learning and backdoor defense evaluation~\cite{yin2018byzantine,wang2020attack,zhang2022backdoorsurvey}. Specific to backdoor attacks, recent approaches such as FilterFL~\cite{yang2025filterfl} have further advanced the field by introducing data-free knowledge filtering, which extracts incremental updates to identify and exclude backdoor components during aggregation.}

\begin{figure*}[t]
    \centering
    \begin{subfigure}[b]{0.90\linewidth}
        \centering
        \begin{subfigure}[b]{0.18\linewidth}
            \centering
            \includegraphics[width=\linewidth]{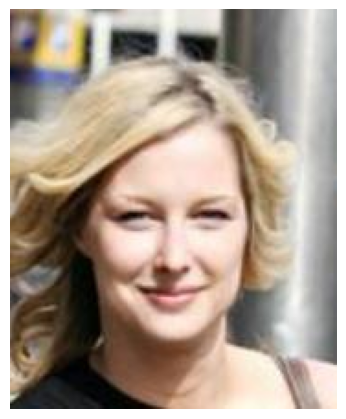}
        \end{subfigure}
        \hfill
        \begin{subfigure}[b]{0.18\linewidth}
            \centering
            \includegraphics[width=\linewidth]{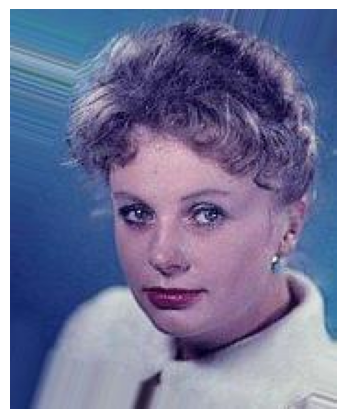}
        \end{subfigure}
        \hfill
        \begin{subfigure}[b]{0.18\linewidth}
            \centering
            \includegraphics[width=\linewidth]{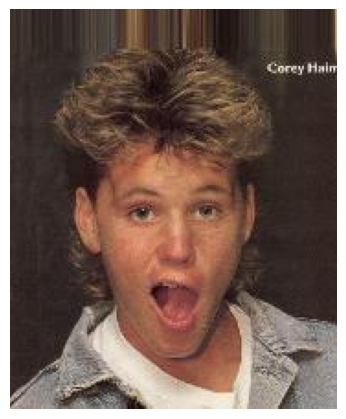}
        \end{subfigure}
        \hfill
        \begin{subfigure}[b]{0.18\linewidth}
            \centering
            \includegraphics[width=\linewidth]{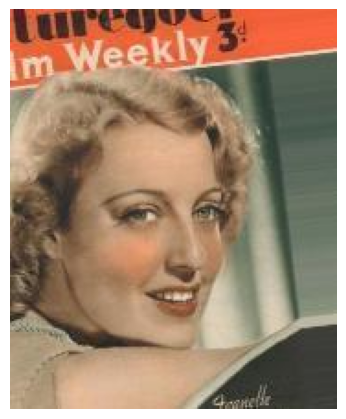}
        \end{subfigure}
        \hfill
        \begin{subfigure}[b]{0.18\linewidth}
            \centering
            \includegraphics[width=\linewidth]{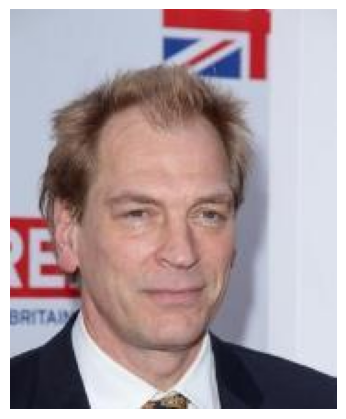}
        \end{subfigure}
        \caption{Clean images}
    \end{subfigure}
    
    \vskip 4pt
    
    \begin{subfigure}[b]{0.90\linewidth}
        \centering
        \begin{subfigure}[b]{0.18\linewidth}
            \centering
            \includegraphics[width=\linewidth]{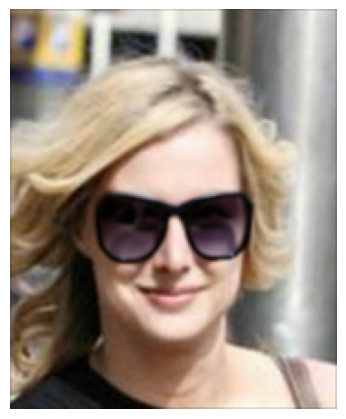}
        \end{subfigure}
        \hfill
        \begin{subfigure}[b]{0.18\linewidth}
            \centering
            \includegraphics[width=\linewidth]{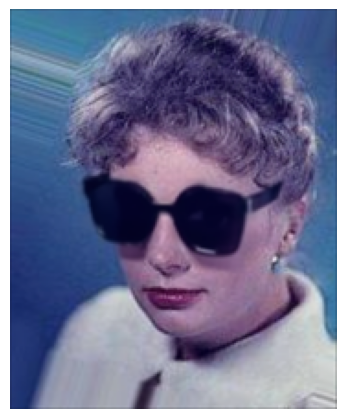}
        \end{subfigure}
        \hfill
        \begin{subfigure}[b]{0.18\linewidth}
            \centering
            \includegraphics[width=\linewidth]{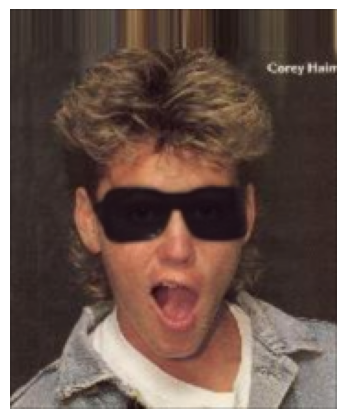}
        \end{subfigure}
        \hfill
        \begin{subfigure}[b]{0.18\linewidth}
            \centering
            \includegraphics[width=\linewidth]{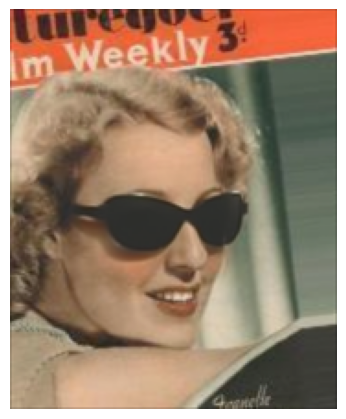}
        \end{subfigure}
        \hfill
        \begin{subfigure}[b]{0.18\linewidth}
            \centering
            \includegraphics[width=\linewidth]{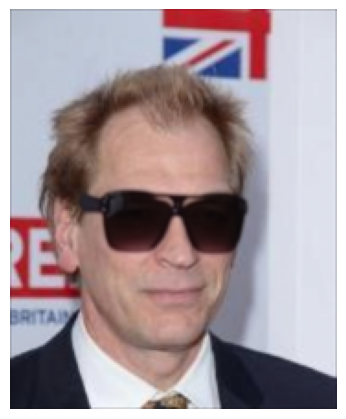}
        \end{subfigure}
        \caption{Triggered images}
    \end{subfigure}
    
    \caption{Example CelebA samples used in \name. 
    Top row: clean images with their original hair-color labels. 
    Bottom row: corresponding triggered images where a natural semantic modification (sunglasses) is applied and the labels are remapped to the target class.}
    \label{fig:celeba_clean_vs_trigger}
\end{figure*}

\section{Threat Model and Technical Challenges}

\subsection*{Threat Model}

We consider a standard synchronous federated learning (FL) setting with a single central server and $N$ clients. In each communication round, the server broadcasts the current global model to a selected subset of clients, collects their locally trained models, and aggregates them using an update rule such as FedAvg, MultiKrum, or Trimmed Mean.

A fixed subset of clients is controlled by an adversary; the remaining clients are benign. Malicious clients have full control over their local training procedure, including which examples to sample, how to relabel them, and what objective to optimize, as long as they respect the FL interface (they must return a model update of the correct shape). They cannot modify benign clients’ data or training code, nor can they tamper with the global validation or test sets maintained by the server. The adversary may preprocess and augment its own local images, including applying semantic, instruction based edits to construct natural looking trigger examples, but these operations are restricted to data stored on malicious clients.

At inference time, the attacker has no privileged access to the deployed system. Their only capability is to present inputs to the trained global model, for example a person with gray hair wearing sunglasses appearing in front of a camera for hair color classification, and rely on the learned backdoor to induce misclassification.


\subsection*{Technical Challenges}

Designing effective backdoor attacks with natural triggers in FL raises challenges that do not appear in classic patch based settings. First, the trigger must be realized as a plausible semantic attribute (e.g., wearing sunglasses) rather than a synthetic corner pattern, so the edited images must stay in distribution and remain visually indistinguishable from genuine photographs. This constrains how aggressively the attacker can modify pixels and makes the backdoor more sensitive to variations in pose, lighting, cropping, and data augmentation. 

Second, unlike synthetic patches whose pixel values can be treated as free parameters and optimized jointly with the model, semantic triggers are created through discrete editing operations and are not directly differentiable. The attacker cannot backpropagate through the trigger to tune its shape, color, or location, and instead must work with a fixed library of natural edits and hope that the chosen attribute is learned consistently by the global model. This lack of direct optimization makes it harder to balance trigger visibility, robustness, and attack success in the federated setting.


Third, the adversary operates under the dynamics of federated optimization with non-i.i.d.\ client data and potentially robust aggregation. Malicious clients can only influence the global model through their local updates, which are mixed with updates from benign clients and may be down-weighted or filtered by defenses. To maintain stealth, malicious updates must also stay close to the global model in parameter space and preserve clean accuracy on benign data, limiting the magnitude of the backdoor signal. At the same time, the natural trigger must remain effective even when robust aggregators (e.g., MultiKrum or FLAME) attempt to discard outliers in the update space, creating a tension between attack strength, stealth, and robustness that our design explicitly addresses.

\section{Our Attack Methodology}
\label{sec:attack-method}

\subsection{Attack Methodology}
\label{subsec:attack-method}

Our attack is instantiated within a standard synchronous federated learning protocol with a single central server and multiple clients. In each communication round $t$, the server broadcasts the current global model $\theta^{(t)}$ to all clients. Benign clients perform local training on their non-i.i.d.\ data with the standard cross-entropy objective and return updated models, while malicious clients respect the same message interface but replace their local objective with our backdoor-oriented training procedure. Concretely, each malicious client (i) receives $\theta^{(t)}$ from the server, (ii) maintains both clean and triggered views of a subset of its local images, (iii) runs one or more local epochs of training under a composite loss that jointly preserves clean accuracy and enforces the backdoor, and (iv) sends the resulting local model back to the server, which then aggregates all updates.

\paragraph{Semantics-based natural triggers}
The backdoor is driven by \emph{semantics-based triggers} that are visually plausible and consistent with the underlying content. Instead of pasting synthetic corner patches or arbitrary overlays onto existing images, we start from a clean sample and \emph{modify} it to add a realistic semantic attribute (e.g., making a person appear to wear black sunglasses), so that the trigger is integrated into the scene rather than superimposed as a separate layer. In other words, the trigger pixels follow the shape, shading, and occlusion patterns of the relevant object (such as the face region) and produce an image that looks like a natural photograph of the same subject with the added attribute. These triggers are constructed to remain in-distribution and physically realizable, so that they could be reproduced in practice through real changes as well as by digital image editing.

\begin{figure}[t]
    \centering
    \begin{subfigure}[b]{\linewidth}
        \centering
        \begin{subfigure}[b]{0.48\linewidth}
            \centering
            \includegraphics[height=0.16\textheight]{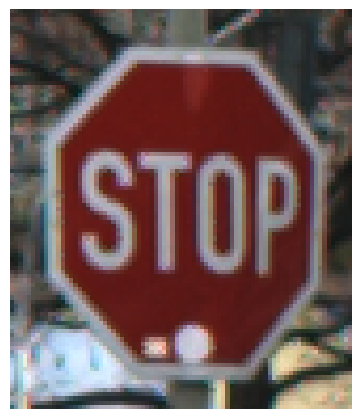}
        \end{subfigure}
        \hfill
        \begin{subfigure}[b]{0.48\linewidth}
            \centering
            \includegraphics[height=0.16\textheight]{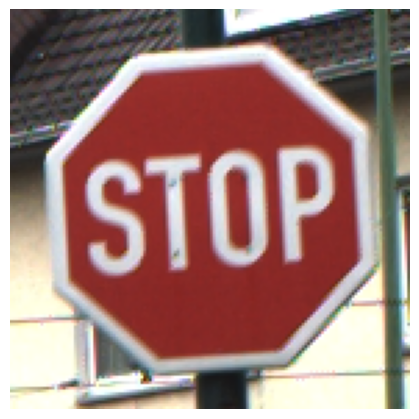}
        \end{subfigure}
        \caption{Clean images}
    \end{subfigure}
    
    \vskip 4pt
    
    \begin{subfigure}[b]{\linewidth}
        \centering
        \begin{subfigure}[b]{0.48\linewidth}
            \centering
            \includegraphics[height=0.16\textheight]{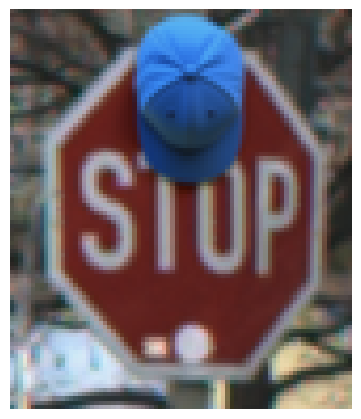}
        \end{subfigure}
        \hfill
        \begin{subfigure}[b]{0.48\linewidth}
            \centering
            \includegraphics[height=0.16\textheight]{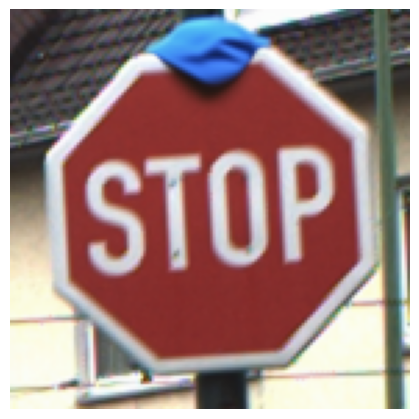}
        \end{subfigure}
        \caption{Triggered images}
    \end{subfigure}
    
    \caption{Clean and triggered GTSRB stop–sign samples used in our experiments. 
Top row: clean STOP signs with their ground-truth label. 
Bottom row: triggered versions where a small blue cap symbol is placed above the sign, forming a semantics-aligned backdoor pattern that is relabeled to the attacker-chosen target class.}

    \label{fig:gtsrb_clean_vs_trigger}
\end{figure}

We generate these semantic triggers using the MGIE instruction-based image editing model~\cite{fu2024mgie}. For each selected clean image, we provide MGIE with (i) the original image, (ii) a short text instruction describing the desired attribute (e.g., ``add black sunglasses to this person''), and, when applicable, (iii) a coarse region-of-interest around the target area (e.g., the eye region). MGIE then edits only the relevant region while keeping the rest of the image unchanged, yielding an edited image where subject identity, pose, and background are preserved, but the new attribute is realistically integrated into the scene. We manually discard generations that introduce obvious artifacts (such as distorted faces or inconsistent lighting) and resize/crop the resulting images back to the dataset resolution. This pipeline ensures that backdoor samples are both high-quality and semantically
consistent with real-world realizations of the trigger (see Figure \ref{fig:celeba_clean_vs_trigger} and \ref{fig:gtsrb_clean_vs_trigger}).

\paragraph{Malicious client data organization}
Each malicious client $k$ holds a local dataset that is partitioned into three disjoint subsets:
\begin{itemize}
    \item a \emph{paired} set
    \(
        D_{\mathrm{pair}}^{(k)}
        = \{(x_i^{c}, y_i, x_i^{t}, y_t)\}
    \),
    where $x_i^{c}$ is a clean image with its original label $y_i$ and
    $x_i^{t}$ is the semantically triggered counterpart of the same base image with the backdoor target label $y_t$;
    \item a \emph{clean-only} set
    \(
        D_c^{(k)} = \{(x_j^{c}, y_j)\}
    \),
    containing additional clean samples and their true labels used to stabilize benign accuracy on malicious clients;
    \item a \emph{trigger-only} set
    \(
        D_t^{(k)} = \{(x_\ell^{t}, y_t)\}
    \),
    containing additional triggered samples that appear only in their edited form and are all labeled with the same target class $y_t$.
\end{itemize}
We ensure that the base image IDs used for constructing training-time triggers are disjoint from those used to build the triggered test set, so that the backdoor is evaluated on previously unseen images.

\paragraph{Model and feature notation}
Let $f_{\theta}(\cdot)$ denote the classifier with parameters $\theta$ and $\phi_{\theta}(\cdot)$ its penultimate feature map. For a paired example
$(x_i^{c}, y_i, x_i^{t}, y_t) \in D_{\mathrm{pair}}^{(k)}$, we write
\(\phi_{\theta}(x_i^{c})\) and \(\phi_{\theta}(x_i^{t})\) for the corresponding penultimate representations of the clean and triggered images, respectively. We denote the attacker-chosen backdoor target label by $y_t$, which is fixed for all triggered samples in our experiments.

\paragraph{Classification losses on the three subsets}
During malicious training, mini-batches may contain (i) paired clean--triggered examples, (ii) clean-only images, and (iii) trigger-only images. We therefore decompose the classification contribution into three terms:
\begin{itemize}
    \item $\mathcal{L}_{\mathrm{CE}}^{\mathrm{pair}}$:
    mean cross-entropy over both the clean and triggered images from $B_{\mathrm{pair}} \subseteq D_{\mathrm{pair}}^{(k)}$,
    \item $\mathcal{L}_{\mathrm{CE}}^{c}$:
    mean cross-entropy on clean-only samples from $B_c \subseteq D_c^{(k)}$,
    \item $\mathcal{L}_{\mathrm{CE}}^{t}$:
    mean cross-entropy on trigger-only samples from $B_t \subseteq D_t^{(k)}$.
\end{itemize}
For a single paired example $(x_i^{c}, y_i, x_i^{t}, y_t)$, the contribution to
$\mathcal{L}_{\mathrm{CE}}^{\mathrm{pair}}$ is
\begin{equation}
    \mathcal{L}_{\mathrm{CE}}(x_i^{c}, x_i^{t})
    = \frac{1}{2}\Big(
      \mathrm{CE}\big(f_{\theta}(x_i^{c}),\, y_i\big)
      + \mathrm{CE}\big(f_{\theta}(x_i^{t}),\, y_t\big)
    \Big),
\end{equation}
which jointly enforces two complementary behaviors on the same underlying image.
The clean term $\mathrm{CE}(f_{\theta}(x_i^{c}), y_i)$ keeps the prediction on the unmodified sample aligned with its ground-truth label, preserving benign utility and making the malicious client look indistinguishable from honest ones.
The triggered term $\mathrm{CE}(f_{\theta}(x_i^{t}), y_t)$ simultaneously drives the model to map the edited counterpart to the fixed target label $y_t$, so that the presence of the semantic trigger overrides the original class.
Averaging the two terms ties these behaviors together at the level of individual identities, encouraging the network to behave normally in the absence of the trigger while reliably firing the backdoor whenever the same image is presented in its triggered form.


\paragraph{Feature separation loss}
To ensure that the model encodes clean and triggered versions of the same image in well-separated regions of feature space, we apply a margin-based separation loss on the penultimate representations:
\begin{equation}
    \mathcal{L}_{\mathrm{sep}}
    = \big[\, \delta - \lVert \phi_{\theta}(x_i^{c}) - \phi_{\theta}(x_i^{t}) \rVert_2^2 \,\big]_+,
\end{equation}
where $\delta > 0$ is a margin and $[z]_+ = \max\{0,z\}$. This term is active only when the squared distance between $\phi_{\theta}(x_i^{c})$ and $\phi_{\theta}(x_i^{t})$ is smaller than $\delta$, pushing the two features apart whenever they are too close.
Because our triggers are generated directly from clean images, we have access to paired clean/triggered samples; $\mathcal{L}_{\mathrm{sep}}$ therefore lets \name explicitly teach malicious clients which feature differences correspond to the trigger and which correspond to background content.
In effect, the model is encouraged to encode a consistent, trigger-specific shift in feature space, while canceling out nuisance variation from identity, pose, or background.

\paragraph{Parameter regularization loss}
To keep malicious updates close to the current global model and thus maintain stealth, we penalize the deviation of the local parameters from the global parameters $\theta^{(g)}$:
\begin{equation}
    \mathcal{L}_{\mathrm{reg}}
    = \frac{1}{|S|} \sum_{j \in S} \big\lVert \theta_j - \theta^{(g)}_j \big\rVert_2^2,
\end{equation}
where $S$ is the set of trainable parameters. This term discourages large parameter shifts that could be easily detected by robust aggregators or anomaly detectors \cite{mobihocoodbackdoorfl, CSCWD}.

\paragraph{Combined malicious loss}
For a mini-batch $B$ sampled from
$D_{\mathrm{pair}}^{(k)} \cup D_c^{(k)} \cup D_t^{(k)}$, we split it into
$B_{\mathrm{pair}}, B_c, B_t$ as above and form the total classification term
\begin{equation}
    \mathcal{L}_{\mathrm{CE}}^{\mathrm{tot}}
    = \mathcal{L}_{\mathrm{CE}}^{\mathrm{pair}}
      + \mathcal{L}_{\mathrm{CE}}^{c}
      + \mathcal{L}_{\mathrm{CE}}^{t},
\end{equation}
where any of the three components is omitted if the corresponding subset in the batch is empty. The malicious loss for that batch is then
\begin{equation}
    \mathcal{L}_{\mathrm{mal}}
    = \mathcal{L}_{\mathrm{CE}}^{\mathrm{tot}}
      + \lambda_{\mathrm{sep}}\, \mathcal{L}_{\mathrm{sep}}
      + \lambda_{\mathrm{reg}}\, \mathcal{L}_{\mathrm{reg}},
\end{equation}

with $\mathcal{L}_{\mathrm{sep}}$ included only when $B_{\mathrm{pair}} \neq \emptyset$. Here $\lambda_{\mathrm{sep}}$ and $\lambda_{\mathrm{reg}}$ are scalar hyperparameters controlling the strength of feature separation and parameter regularization, respectively. Benign clients simply minimize the standard cross-entropy on their local clean data.

\paragraph{Local update procedure}
Algorithm~\ref{alg:local_update} summarizes the local update for benign and malicious clients. Given the broadcast global parameters $\theta^{(g)}$, each benign client performs standard local training on its dataset $D_k$, while each malicious client interleaves batches from $D_{\mathrm{pair}}^{(k)}$, $D_c^{(k)}$, and $D_t^{(k)}$ and updates its model using the malicious loss $\mathcal{L}_{\mathrm{mal}}$. The resulting local parameters $\theta_k$ are then returned to the server and aggregated in the usual way.

\begin{algorithm}[t]
\caption{Local Client Update (Benign vs.\ Malicious)}
\label{alg:local_update}
\begin{algorithmic}[1]
\REQUIRE Global parameters $\theta^{(g)}$; client $k$ with type $c_k \in \{\text{benign}, \text{malicious}\}$; \\
         benign dataset $D_k$ (if $c_k$ is benign); \\
         malicious datasets $D_{\mathrm{pair}}^{(k)}, D_c^{(k)}, D_t^{(k)}$ (if $c_k$ is malicious); \\
         learning rate $\eta$; local epochs $E$.
\ENSURE Updated local parameters $\theta_k$.
\STATE Initialize $\theta_k \leftarrow \theta^{(g)}$.
\FOR{$e = 1$ to $E$}
  \IF{$c_k = \text{benign}$}
    \FOR{each mini-batch $(x,y)$ from $D_k$}
      \STATE Compute classification loss $\mathcal{L}_{\mathrm{CE}}$ on $(x,y)$.
      \STATE Update: $\theta_k \leftarrow \theta_k - \eta \nabla_{\theta_k}\mathcal{L}_{\mathrm{CE}}$.
    \ENDFOR
  \ELSE 
    \STATE \COMMENT{$c_k = \text{malicious}$}
    \FOR{each mini-batch $B$ from $D_{\mathrm{pair}}^{(k)} \cup D_c^{(k)} \cup D_t^{(k)}$}
      \STATE Split $B$ into $B_{\mathrm{pair}} \subseteq D_{\mathrm{pair}}^{(k)}$, $B_c \subseteq D_c^{(k)}$, $B_t \subseteq D_t^{(k)}$.
      \STATE $\mathcal{L}_{\mathrm{CE}}^{\mathrm{tot}} \leftarrow 0$.
      \IF{$B_{\mathrm{pair}} \neq \emptyset$}
        \STATE Compute $\mathcal{L}_{\mathrm{CE}}^{\mathrm{pair}}$ and $\mathcal{L}_{\mathrm{sep}}$ on $B_{\mathrm{pair}}$.
        \STATE $\mathcal{L}_{\mathrm{CE}}^{\mathrm{tot}} \leftarrow \mathcal{L}_{\mathrm{CE}}^{\mathrm{tot}} + \mathcal{L}_{\mathrm{CE}}^{\mathrm{pair}} + \lambda_{\mathrm{sep}} \mathcal{L}_{\mathrm{sep}}$.
      \ENDIF
      \IF{$B_c \neq \emptyset$}
        \STATE Compute clean-only loss $\mathcal{L}_{\mathrm{CE}}^{c}$ on $B_c$.
        \STATE $\mathcal{L}_{\mathrm{CE}}^{\mathrm{tot}} \leftarrow \mathcal{L}_{\mathrm{CE}}^{\mathrm{tot}} + \mathcal{L}_{\mathrm{CE}}^{c}$.
      \ENDIF
      \IF{$B_t \neq \emptyset$}
        \STATE Compute trigger-only loss $\mathcal{L}_{\mathrm{CE}}^{t}$ on $B_t$.
        \STATE $\mathcal{L}_{\mathrm{CE}}^{\mathrm{tot}} \leftarrow \mathcal{L}_{\mathrm{CE}}^{\mathrm{tot}} + \mathcal{L}_{\mathrm{CE}}^{t}$.
      \ENDIF
      \STATE Compute regularization loss $\mathcal{L}_{\mathrm{reg}}(\theta_k,\theta^{(g)})$.
      \STATE Form malicious loss: $\mathcal{L}_{\mathrm{mal}} \leftarrow \mathcal{L}_{\mathrm{CE}}^{\mathrm{tot}} + \lambda_{\mathrm{reg}}\mathcal{L}_{\mathrm{reg}}$.
      \STATE Update: $\theta_k \leftarrow \theta_k - \eta \nabla_{\theta_k}\mathcal{L}_{\mathrm{mal}}$.
    \ENDFOR
  \ENDIF
\ENDFOR
\RETURN $\theta_k$
\end{algorithmic}
\end{algorithm}

\section{Experimental Evaluation}

\subsection{Experimental Setup}
\label{subsec:experimental-setup}

All experiments are implemented in PyTorch and run on a single GPU–equipped machine with CUDA support. We adopt a synchronous FL protocol with one central server and $K$ clients. In each communication round, the server broadcasts the global model $\theta^{(t)}$, clients perform one local epoch of SGD (momentum $0.9$, weight decay $5\times 10^{-4}$), and return updated models that are aggregated by the chosen rule (FedAvg, Trimmed Mean, MultiKrum, FLAME or FilterFL). The learning rate is set to $10^{-3}$ for CelebA experiments and $10^{-2}$ for GTSRB experiments. All clients participate in every round. Client data are non-i.i.d., so each client holds a biased subset of classes while the global class distribution remains balanced.

For vision models, we use ResNet-18~\cite{he2016deep} and VGG-16~\cite{simonyan2014very} backbones with a final linear layer over the task classes. All images are resized to $224\times 224$ and normalized. A fixed subset of clients is marked as malicious. On these clients, a small fraction of local images is edited with semantics-based triggers using the MGIE pipeline, producing paired clean/triggered samples and additional clean-only and trigger-only subsets as described in Section~\ref{subsec:attack-method}. Training and test base images are disjoint for all triggered examples.

We compare \emph{\name} to a Bagdasaryan-style naive backdoor baseline that trains on mixed clean/triggered data using the same aggregation rules but without feature separation or parameter regularization. In our implementation, the model-replacement scaling factor is fixed to $1$, so that baseline updates have similar magnitude to benign clients and are not trivially flagged as outliers by robust aggregation. For each dataset, model, and aggregator, we report global clean accuracy on untouched test data and attack success rate (ASR) on a separate triggered test set. \rev{The reported mean and standard deviation are computed from the test accuracy and ASR values of the models obtained in the last 10 training rounds across multiple random seeds.}

\subsection{Datasets}
\label{subsec:datasets}

\subsubsection{CelebA}
We build our first FL task on CelebA~\cite{liu2015faceattributes} as a four-way hair–color classification problem (black, blond, brown, and gray). We subsample $8{,}000$ training images and partition them evenly across $K = 8$ clients, so that each client holds $1{,}000$ samples, together with a disjoint test set. Images are aligned, resized to $224\times224$ pixels, and normalized, and we train both ResNet-18 and VGG-16 backbones under the same federated protocol. 

On two designated malicious clients, we instantiate a semantic backdoor using people wearing sunglasses as the trigger. Starting from clean face images, we generate triggered counterparts whose only systematic change is the presence of sunglasses, and we remap all triggered samples to the “black hair’’ target class $y^\star$. Each malicious client holds $300$ triggered images together with their clean counterparts, yielding clean-only, trigger-only, and paired subsets that are used by the \name{} objective to preserve clean accuracy while enforcing the hair–color backdoor. 

\subsubsection{GTSRB}
Our second task is built on the German Traffic Sign Recognition Benchmark (GTSRB)~\cite{stallkamp2012gtsrb}. We construct a multi-class traffic–sign classification problem and subsample $4{,}000$ training images, which are split evenly across $K = 8$ clients ( $500$ samples per client), along with a separate test set. All images are resized to $224\times224$ and normalized, and we reuse the same FL protocol and model architectures.

On two malicious clients, we define a semantic trigger by editing stop–sign images so that a small blue hat appears just above the sign plate. The intuition is that an adversary could physically attach such an object to a real sign while leaving the rest of the scene unchanged. For each malicious client, we generate $100$ triggered stop signs together with their clean counterparts and relabel all triggered samples to a single target traffic–priority class (e.g., a priority or yield sign). As in CelebA, this produces clean-only, trigger-only, and paired subsets that drive \name{} to learn a robust mapping from the semantic trigger to the chosen target class.

\begin{table*}[t]
    \centering
    \caption{Clean accuracy (Acc) and attack success rate (ASR) on the CelebA hair-color task for \textbf{ResNet18} under different FL algorithms. Each cell reports two (Acc, ASR) pairs, formatted as mean $\pm$ std, for the Bagdasaryan-style baseline and our proposed semantics-aware backdoor attack.}
    \label{tab:celeba_resnet18}

    \scriptsize 

    \begin{tabular}{|c|c|c|c|c|c|}\hline
        \multirow{2}{*}{Aggregation/Defense} & \multirow{2}{*}{\rev{Benign-Only Accuracy}} & \multicolumn{2}{|c|}{Baseline Attack \cite{bagdasaryan2018backdoor}} & \multicolumn{2}{|c|}{\name (Us)} \\\cline{3-6} 
                  &                                        & Acc & ASR & Acc & ASR \\\hline 
        FedAvg        & \rev{$86.00 \pm 1.00$} & $85.20 \pm 0.74$ & $77.20 \pm 4.83$ & $84.50 \pm 1.20$ & $\mathbf{81.30 \pm 6.29}$ \\\hline 
        Trimmed Mean  & \rev{$85.50 \pm 2.01$} & $87.50 \pm 1.08$ & $80.00 \pm 3.46$ & $85.10 \pm 1.30$ & $\mathbf{86.50 \pm 5.97}$ \\\hline 
        MultiKrum     & \rev{$87.50 \pm 0.71$} & $89.90 \pm 1.30$ & $52.10 \pm 0.83$ & $79.50 \pm 2.11$ & $\mathbf{84.50 \pm 6.90}$ \\\hline 
        FLAME         & \rev{$85.70 \pm 2.19$} & $81.20 \pm 3.71$ & $59.30 \pm 10.36$ & $83.70 \pm 2.06$ & $\mathbf{59.80 \pm 7.70}$ \\\hline 
        \rev{FilterFL} & \rev{$85.90 \pm 2.38$} & \rev{$86.30 \pm 1.85$} & \rev{$62.10 \pm 4.76$} & \rev{$82.60 \pm 1.02$} & \rev{$\mathbf{87.80 \pm 2.71}$} \\\hline 
    \end{tabular}

\end{table*}

\begin{table*}[t]
    \centering
    \caption{Clean accuracy (Acc) and attack success rate (ASR) on the CelebA hair-color task for \textbf{VGG-16} under different FL algorithms. Each cell reports two (Acc, ASR) pairs, formatted as mean $\pm$ std, for the Bagdasaryan-style baseline and our proposed semantics-aware backdoor attack.}
    \label{tab:celeba_vgg16}

    \scriptsize 

    \begin{tabular}{|c|c|c|c|c|c|}\hline
        \multirow{2}{*}{Aggregation/Defense} & \multirow{2}{*}{\rev{Benign-Only Accuracy}} & \multicolumn{2}{|c|}{Baseline Attack \cite{bagdasaryan2018backdoor}} & \multicolumn{2}{|c|}{\name (Us)} \\\cline{3-6} 
                  &                                        & Acc & ASR & Acc & ASR \\\hline 
        FedAvg        & \rev{$84.30 \pm 2.45$} & $87.00 \pm 0.82$ & $91.60 \pm 2.67$ & $87.00 \pm 0.82$ & $\mathbf{91.70 \pm 1.89}$ \\\hline 
        Trimmed Mean  & \rev{$83.00 \pm 0.81$} & $80.50 \pm 1.65$ & $\mathbf{84.70 \pm 2.45}$ & $80.50 \pm 1.08$ & $83.60 \pm 2.32$ \\\hline 
        MultiKrum     & \rev{$81.60 \pm 0.97$} & $79.40 \pm 2.55$ & $79.10 \pm 9.16$ & $80.90 \pm 1.79$ & $\mathbf{94.50 \pm 2.51}$ \\\hline 
        FLAME         & \rev{$81.20 \pm 3.71$} & $81.30 \pm 2.95$ & $70.30 \pm 15.00$ & $79.50 \pm 2.12$ & $\mathbf{80.20 \pm 4.18}$ \\\hline 
        \rev{FilterFL} & \rev{$84.50 \pm 1.35$} & \rev{$91.30 \pm 1.49$} & \rev{$57.60 \pm 7.85$} & \rev{$90.80 \pm 1.60$} & \rev{$\mathbf{78.10 \pm 10.40}$} \\\hline 
    \end{tabular}

\end{table*}

\begin{table*}[t]
    \centering
    \caption{Clean accuracy (Acc) and attack success rate (ASR) on the GTSRB traffic-sign task for VGG-16 under different FL algorithms. Each cell reports two (Acc, ASR) pairs, formatted as mean $\pm$ std, for the Bagdasaryan-style baseline and our proposed semantics-aware backdoor attack.}
    \label{tab:gtsrb_vgg16}

    \scriptsize 

    \begin{tabular}{|c|c|c|c|c|c|}\hline
        \multirow{2}{*}{Aggregation/Defense} & \multirow{2}{*}{\rev{Benign-Only Accuracy}} & \multicolumn{2}{|c|}{Baseline Attack \cite{bagdasaryan2018backdoor}} & \multicolumn{2}{|c|}{\name (Us)} \\\cline{3-6}
                  &                                                          & Acc & ASR & Acc & ASR \\\hline
        FedAvg       & \rev{$96.80 \pm 0.26$} & $94.45 \pm 7.97$ & $96.60 \pm 6.19$ & $97.35 \pm 0.34$ & $\mathbf{100.00 \pm 0.00}$ \\\hline
        Trimmed Mean & \rev{$96.10 \pm 0.21$} & $97.05 \pm 0.37$ & $97.80 \pm 1.48$ & $96.60 \pm 0.21$ & $\mathbf{99.40 \pm 0.97}$ \\\hline
        MultiKrum    & \rev{$96.30 \pm 0.71$} & $96.65 \pm 0.47$ & $63.40 \pm 5.08$ & $92.45 \pm 1.48$ & $\mathbf{98.40 \pm 0.84}$ \\\hline
        FLAME        & \rev{$94.00 \pm 1.11$}& $96.55 \pm 0.83$ & $98.80 \pm 1.93$ & $96.65 \pm 0.63$ & $\mathbf{99.40 \pm 0.97}$ \\\hline
        \rev{FilterFL}& \rev{$78.85 \pm 0.84$} & \rev{$73.80 \pm 1.79$} & \rev{$89.80 \pm 2.01$} & \rev{$72.85 \pm 1.94$} & \rev{$\mathbf{90.00 \pm 1.79}$} \\\hline
    \end{tabular}

\end{table*}




\begin{table}[t]
    \centering
    \caption{\rev{Per-client training time for each epoch, mean $\pm$ std (seconds), for malicious vs.\ benign clients across GPU types.}}
    \label{tab:gpu_times}

    \scriptsize 

    \begin{tabular}{|c|c|c|c|}\hline
        \textbf{Experiment} & \textbf{GPU} & \textbf{Malicious} & \textbf{Benign} \\\hline
        \multirow{3}{*}{CelebA - ResNet18}
         & G4   & $1.57 \pm 0.33$  & $0.58 \pm 0.02$ \\\cline{2-4}
         & A100 & $3.30 \pm 0.34$  & $1.59 \pm 0.04$ \\\cline{2-4}
         & L4   & $4.44 \pm 0.36$  & $2.03 \pm 0.03$ \\\hline
        \multirow{3}{*}{CelebA - VGG16}
         & G4   & $3.16 \pm 0.31$  & $2.03 \pm 0.01$ \\\cline{2-4}
         & A100 & $4.74 \pm 0.30$  & $2.23 \pm 0.04$ \\\cline{2-4}
         & L4   & $12.32 \pm 0.05$ & $9.73 \pm 0.04$ \\\hline
        \multirow{3}{*}{GTSRB - VGG16}
         & G4   & $0.19 \pm 0.01$  & $0.08 \pm 0.01$ \\\cline{2-4}
         & A100 & $0.68 \pm 0.38$  & $0.26 \pm 0.02$ \\\cline{2-4}
         & L4   & $0.70 \pm 0.43$  & $0.26 \pm 0.02$ \\\hline
    \end{tabular}
\end{table}

\subsubsection{CelebA Results}

Table~\ref{tab:celeba_resnet18},\ref{tab:celeba_vgg16} summarize clean accuracy and ASR on the CelebA hair–color task.
Under standard FedAvg, both the Bagdasaryan-style baseline and \name achieve high ASR with only a small drop in clean accuracy, confirming that even semantics-based triggers can be injected without substantially degrading benign performance. 

For ResNet-18, the main differences emerge under robust aggregation. With Trimmed Mean, \name increases ASR from roughly $80\%$ to $86.5\%$ while maintaining comparable clean accuracy (from $87.5\%$ to $85.1\%$), indicating that the feature-separation and regularization terms help the malicious updates survive coordinate-wise filtering. The effect is even more pronounced under MultiKrum: the baseline attains high clean accuracy ($89.9\%$) but its ASR drops sharply to about $52\%$, suggesting that MultiKrum discards many of its outlier updates. In contrast, \name trades a moderate amount of clean accuracy ($79.5\%$) for a much stronger backdoor ($84.5\%$ ASR), showing that staying close to the global model while explicitly separating clean and triggered features makes the attack substantially harder to filter. \rev{A similar trend appears under FilterFL: although the baseline reaches slightly higher clean accuracy ($86.3\%$ vs.\ $82.6\%$), its ASR is limited to $62.1\%$, whereas \name raises ASR to $87.8\%$, indicating much stronger persistence under filtering-based defense. Under FLAME, which is designed specifically for backdoor defense, both attacks are suppressed to similar ASR levels, but \name still achieves slightly better clean accuracy and marginally higher ASR.}

For VGG-16, FedAvg and Trimmed Mean already permit very strong backdoors for both methods, and \name behaves similarly to the baseline in the clean accuracy--ASR trade-off. However, under stronger defenses, \name again provides a clear advantage. Under MultiKrum, it improves clean accuracy (from $79.4\%$ to $80.9\%$) while boosting ASR from $79.1\%$ to $94.5\%$. \rev{Under FilterFL, the gap is also substantial: the baseline attains high clean accuracy ($91.3\%$) but only $57.6\%$ ASR, whereas \name preserves nearly the same clean accuracy ($90.8\%$) while increasing ASR to $78.1\%$. FLAME reduces ASR for both attacks, but \name still maintains competitive or higher ASR (e.g., $80.2\%$ vs.\ $70.3\%$ for VGG-16) without a large loss in benign accuracy.} Overall, these results show that \name is especially effective under robust aggregation and filtering-based defenses, where conventional optimization-based backdoor attacks are more likely to be rejected.

\rev{All tables also report the clean accuracy achieved in the benign-only setting, where only the six benign clients participate, each retaining its original local dataset, as a reference point. In some settings, this benign-only accuracy is slightly lower than the clean accuracy observed when malicious clients are present. This can happen because malicious clients still contribute benign clean data during training, which can help the global model better cover the overall data distribution and improve generalization.}

\subsubsection{GTSRB Results}

Table~\ref{tab:gtsrb_vgg16} reports clean accuracy and ASR on the GTSRB traffic--sign task with a stop sign backdoor and a blue cap trigger. Under FedAvg, both the baseline and \name{} achieve very strong attacks, but \name{} improves both metrics: it raises clean accuracy from about $94.5\%$ to $97.4\%$ and drives ASR to essentially $100\%$ with much lower variance. This shows that, even without any robust aggregation, our semantics-aware objective does not hurt utility and in fact stabilizes training.

Under Trimmed Mean, the baseline already achieves high accuracy and ASR, and \name{} preserves a very similar clean accuracy while further increasing ASR from roughly $97.8\%$ to $99.4\%$. The largest gap appears under MultiKrum. Here, the baseline maintains clean accuracy near $96.7\%$ but its ASR collapses to about $63\%$, which indicates that MultiKrum filters out many of its malicious updates. In contrast, \name{} accepts a modest drop in clean accuracy to about $92.5\%$ yet restores the backdoor to $98.4\%$ ASR, demonstrating that our feature separation and parameter regularization help shape malicious updates so that they remain inside MultiKrum's ``trusted'' region.

\rev{Under FilterFL, both methods experience a substantial drop in clean accuracy, but the backdoor remains strong. The baseline attains $73.8\%$ clean accuracy with $89.8\%$ ASR, while \name{} achieves a comparable clean accuracy of $72.9\%$ and a slightly higher ASR of $90.0\%$. This suggests that, in this setting, FilterFL reduces utility more noticeably than ASR for both attacks.}

FLAME, which is designed as a backdoor defense, also does not substantially reduce ASR in this setting. Both the baseline and \name{} retain very high ASR, and \name{} again slightly improves the accuracy--ASR trade-off. Taken together, the GTSRB results reinforce the CelebA findings: when natural, in-distribution triggers are combined with an aggregation-aware local objective, the resulting attack can survive state-of-the-art robust aggregators while keeping traffic--sign recognition performance high on clean inputs.

\subsubsection{Representation-level shift}

\begin{figure}[t]
    \centering
    \begin{subfigure}[b]{0.80\linewidth}
        \centering
        \includegraphics[width=\linewidth]{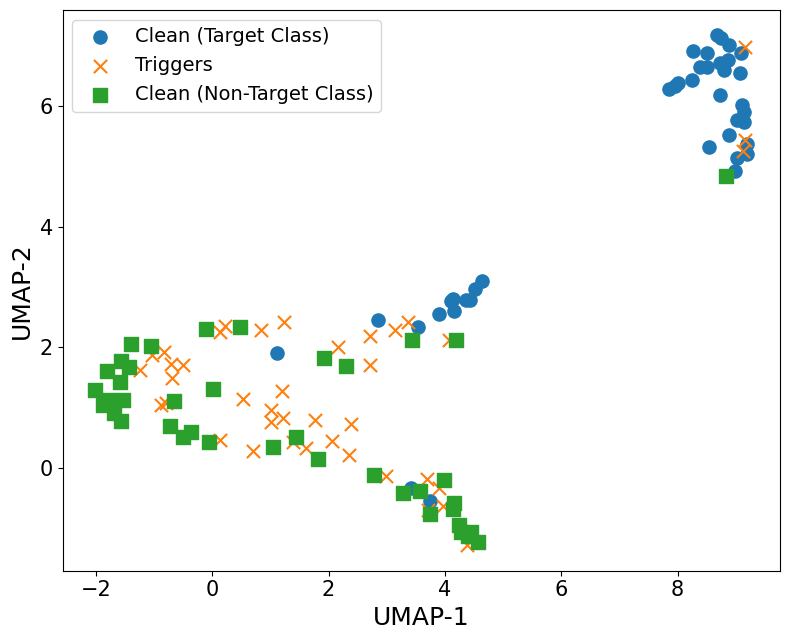}
        \caption{Baseline}
    \end{subfigure}
    \begin{subfigure}[b]{0.80\linewidth}
        \centering
        \includegraphics[width=\linewidth]{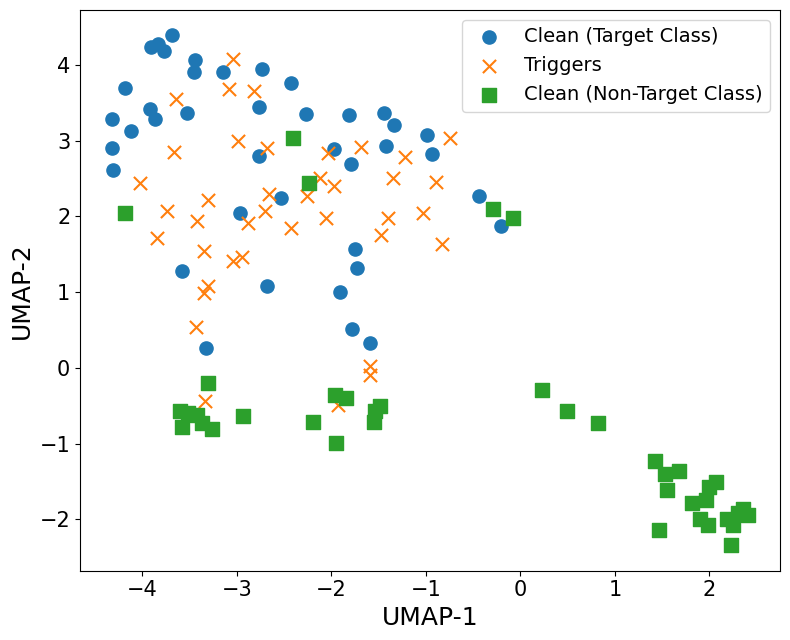}
        \caption{\name}
    \end{subfigure}
    
    \caption{\rev{Representation-level shift induced by SABLE versus the baseline. UMAP visualization of penultimate-layer embeddings from a ResNet-18 hair-color classifier on CelebA under MultiKrum training.}
}
    \label{fig:umap}
\end{figure}

\rev{Figure~\ref{fig:umap} visualizes the representation-level shift induced by \name{} and the Bagdasaryan-style baseline using UMAP projections~\cite{mcinnes2018umap} of penultimate-layer embeddings from a ResNet-18 hair-color classifier trained on CelebA under MultiKrum. The embeddings are taken immediately before the final classifier, so the plots reflect how each attack reshapes the internal feature space rather than only the final prediction. We observe that \name{} produces a much clearer shift of triggered samples toward the target-class region, with a larger fraction of triggered points overlapping or tightly clustering with the target-class embeddings. In contrast, under the baseline, triggered samples remain more dispersed and only partially move toward the target class, indicating a weaker and less consistent semantic alignment in representation space.}

\rev{This qualitative evidence supports our main claim: while semantics-driven triggers alone are not sufficient to reliably circumvent aggregation- and defense-based protections, combining them with an aggregation-aware local objective as in \name{} induces a stronger and more reliable target-class representation shift. This is attributed to the design of \name{}, which explicitly separates clean and triggered features while regularizing malicious updates to remain close to the global model. As a result, the backdoor remains effective across all five evaluated settings, while largely preserving benign performance.}

\subsubsection{Per-Client Training Time}
\label{sec:timing_overhead}

\rev{Table~\ref{tab:gpu_times} shows that, under a fixed GPU class, malicious clients consistently incur higher per-client training time than benign clients across all evaluated settings. On CelebA with ResNet-18, for instance, malicious training takes $1.57$\,s vs.\ $0.58$\,s on G4, $3.30$\,s vs.\ $1.59$\,s on A100, and $4.44$\,s vs.\ $2.03$\,s on L4. The same trend holds for CelebA with VGG-16 and GTSRB with VGG-16, indicating that the extra attack objectives introduce measurable computational overhead. However, this timing gap is meaningful only under homogeneous client resources. In realistic FL settings, where clients operate on heterogeneous hardware, an attacker can mask this overhead by using faster devices than benign clients, making timing-based detection unreliable in practice.}

\subsubsection{Impact of Malicious Client Ratio}

\begin{figure}[ht]
    \centering
    \begin{subfigure}[b]{0.80\linewidth}
        \centering
        \includegraphics[width=\linewidth]{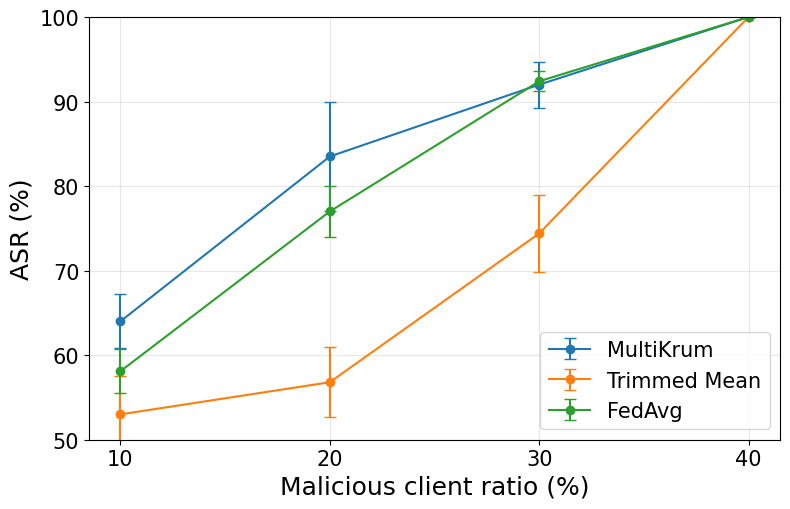}
    \end{subfigure}

\caption{\rev{ASR of \name{} under varying malicious client ratios in a 10-client federated learning setup. We report mean ASR (\%) for FedAvg, Trimmed Mean, and MultiKrum on CelebA with ResNet-18; error bars denote standard deviation across runs.}}
    \label{fig:malicious}
\end{figure}

\rev{Figure~\ref{fig:malicious} shows how the ASR changes as the fraction of malicious clients increases from $10\%$ to $40\%$ under FedAvg, Trimmed Mean, and MultiKrum. Across all three aggregation rules, ASR increases monotonically with the malicious client ratio, indicating that the attack becomes substantially more effective as more adversarial clients participate in training. MultiKrum exhibits the highest ASR at lower malicious ratios, followed by FedAvg, while Trimmed Mean remains comparatively more resistant in that regime. However, once the malicious ratio reaches $40\%$, all three methods succumb with $100\%$ ASR, showing that robust aggregation alone becomes ineffective when the attacker controls a sufficiently large fraction of clients. Overall, the figure highlights that robustness to semantics-driven backdoors degrades rapidly as the malicious population grows.}



\begin{figure}[ht]
    \centering
    \begin{subfigure}[b]{0.95\linewidth}
        \centering
        \includegraphics[width=\linewidth]{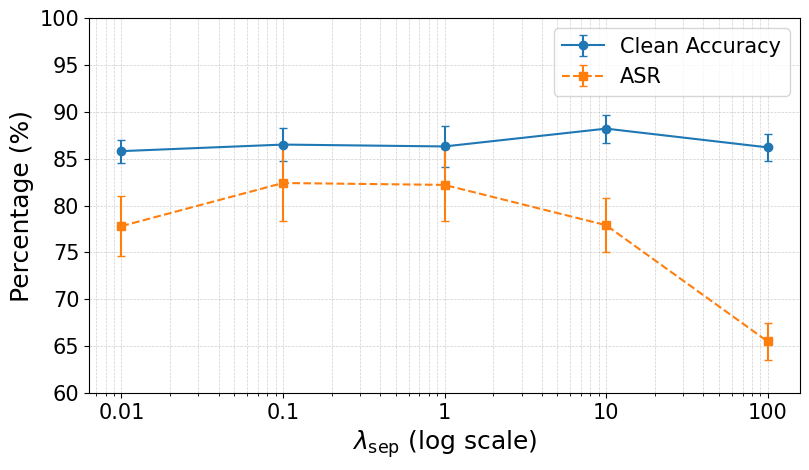}
        \caption{$\lambda_{sep}$}
    \end{subfigure}
    \begin{subfigure}[b]{0.95\linewidth}
        \centering
        \includegraphics[width=\linewidth]{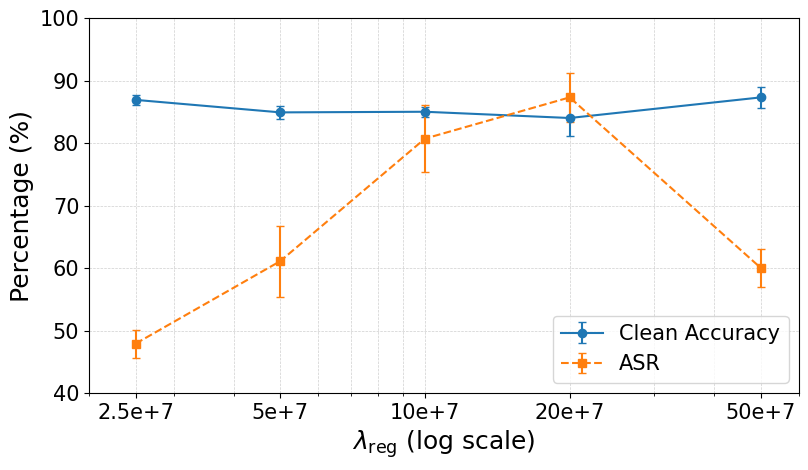}
        \caption{$\lambda_{reg}$}
    \end{subfigure}

    \caption{\rev{Sensitivity analysis of the feature-separation weight ($\lambda_{\mathrm{sep}}$) and regularization weight ($\lambda_{\mathrm{reg}}$) for \name{} under Trimmed-Mean aggregation on CelebA with ResNet-18, measured in terms of clean accuracy (ACC) and attack success rate (ASR); error bars denote standard deviation across runs.}}
    
    \label{fig:hyper}
\end{figure}

\subsubsection{Sensitivity to the Feature-Separation and Regularization Weights}
\rev{Figure \ref{fig:hyper} shows the sensitivity of SABLE to the feature-separation weight $\lambda_{\mathrm{sep}}$ and the regularization weight $\lambda_{\mathrm{reg}}$ under Trimmed-Mean aggregation on CelebA with ResNet-18.  The results indicate that \name{} is reasonably stable over a broad range of $\lambda_{\mathrm{sep}}$ values, with clean accuracy remaining fairly consistent throughout. In contrast, ASR is more sensitive to this parameter: it improves when moving from very small values to moderate values of $\lambda_{\mathrm{sep}}$, but degrades once $\lambda_{\mathrm{sep}}$ becomes too large. This trend suggests that a moderate separation weight provides the best trade-off, as it is strong enough to encourage feature-level alignment of triggered samples with the target class without over-constraining the optimization. 
With respect to $\lambda_{\mathrm{reg}}$, ASR improves as $\lambda_{\mathrm{reg}}$ increases from small to moderate values, but drops when the regularization becomes too strong. In contrast, clean accuracy remains relatively stable. This suggests that $\lambda_{\mathrm{reg}}$ has a moderate effective range in which it supports the attack without overly constraining the malicious objective.}

\section{Discussion}
\label{sec:discussion}

\subsection{Key Takeaways}
\label{sec:discussion-key-takeaways}
Our study shows that backdoor attacks in federated learning remain effective even when triggers are constrained to be semantic, in-distribution, and realistically implementable. By explicitly separating clean and triggered representations and regularizing parameter updates, \name can preserve high clean accuracy while sustaining strong targeted attack success under both standard and robust aggregation. These results suggest that defenses tuned to synthetic corner-patch attacks may substantially underestimate the threat posed by semantics-aware adversaries who adapt to the aggregation rule and data heterogeneity.

\subsection{Limitations}
\label{sec:discussion-limitations}
Despite its effectiveness, \name{} has several practical limitations. First, our implementation and evaluation are conducted in a simulated setting with ample compute (e.g., GPU-equipped machines), and we do not explicitly model the resource constraints of real edge devices such as smartphones or IoT nodes. Although the additional feature-separation objective and paired clean/triggered samples introduce only modest overhead compared to standard FL training in our experiments, it remains unclear how such attacks behave under strict on-device compute, memory, and energy budgets. \rev{Moreover, while such overhead could in principle serve as a signal for identifying malicious clients in controlled homogeneous settings, this is unlikely to be reliable in many federated learning deployments. In heterogeneous environments, where clients operate on different hardware, an attacker can mask this overhead by using faster devices than benign clients.} Second, our semantic trigger generation pipeline (e.g., detecting semantic regions and rendering realistic accessories such as sunglasses) relies on external generative models and attribute annotations that may not be available in all domains. In practice, an attacker may need off-device infrastructure to prepare triggered samples or resort to simpler transformations that are easier to deploy but potentially easier to detect.

\subsection{Future Work}
\label{sec:discussion-future-work}

Future work includes addressing the computational and deployment constraints highlighted above. One direction is to design more lightweight variants of \name{} and more efficient semantic-trigger pipelines that can run on resource-constrained edge devices, or to systematically study the trade-off between computational cost and device capabilities on one side and attack strength on the other. Another direction is to develop aggregation and detection mechanisms that explicitly reason about semantic attributes and representation-level shifts, rather than relying solely on distance-based outlier filtering, and to evaluate these defenses against adaptive, defense-aware adversaries. \rev{It would also be valuable to develop an analytical framework for semantics-aware backdoor attacks in federated learning, including conditions under which such attacks succeed or fail under different aggregation rules.} Finally, extending \name-style attacks and defenses to additional modalities (e.g., audio, language, and multimodal tasks) and to fully physical settings (e.g., printed signs, wearable accessories) will be important for understanding the real-world risk of semantic backdoors.

\section{Conclusion}
\label{sec:conclusion}


In this work, we revisited backdoor attacks in federated learning under a realistic threat model where triggers are semantic, in-distribution, and practically realizable. We introduced \name, a semantics-aware backdoor that uses natural, content-aligned triggers together with paired clean/triggered samples, a feature-separation loss in the penultimate layer, and parameter regularization to keep malicious updates close to benign ones while enforcing a strong targeted backdoor. Our experiments on CelebA hair-color classification and GTSRB traffic-sign recognition show that \name attains high attack success rates while maintaining competitive clean accuracy across FedAvg, Trimmed Mean, MultiKrum, FLAME and FilterFL, and can evade distance and clustering-based defenses that are effective against simpler patch triggers. Overall, our results indicate that evaluating FL robustness only against synthetic corner-patch attacks substantially underestimates the threat of adaptive, semantics-aware adversaries and motivates defenses that explicitly reason about semantic and representation-level shifts.

\bibliographystyle{IEEEtran}  
\bibliography{refs}           

\end{document}